\documentclass[12pt]{article}
\usepackage[cp1251]{inputenc}
\usepackage[russian]{babel}
\begin{document}
\large
\begin{center}
{\bf Neutrino Oscillations in the Scheme of Charge (Couple
Constant) Mixings}  \\
\par
Beshtoev Kh. M.
\par
Joint Institute for Nuclear Research, Joliot Curie 6, 141980
Dubna, Moscow region, Russia \\
\end{center}

\par
{\bf Abstract} \\

\par
In the Standard theory of neutrino oscillations is used scheme of
mass mixings, i.e., oscillation parameters are expressed via terms
of mass matrix. In this work neutrino oscillations generated by
charge (the weak interaction couple constants) mixings are
considered. Expressions for angle mixings and lengths of
oscillations are obtained. The expressions of the probability for
three neutrino oscillations are given. Neutrino oscillations in
this scheme (mechanism) are virtual if neutrino masses are not
equal and real if neutrino masses are equal.

\par
\section{Introduction}

\par
The suggestion that, in analogy with $K^{o},\bar K^{o}$
oscillations, there could be neutrino-antineutrino oscillations (
$\nu \rightarrow \bar \nu$), was considered by Pontecorvo [1] in
1957. It was subsequently considered by Maki et al. [2] and
Pontecorvo [3] that there could be mixings (and oscillations) of
neutrinos of different flavors (i.e., $\nu _{e} \rightarrow \nu
_{\mu }$ transitions).
\par
In the general case there can be two schemes (types) of neutrino
mixings (oscillations): mass mixing schemes and charge  mixings
scheme (as it takes place in the vector dominance model or vector
boson mixings in the standard model of electroweak interactions)
[4].
\par
In the Standard theory of neutrino oscillations [5] is supposed
that physically observed neutrino states $\nu_{e}, \nu_{\mu },
\nu_{\tau}$ have no definite masses and they are directly produced
as mixture of the $\nu_{1}, \nu_{2}, \nu_{3}$ neutrino states. And
if neutrino oscillations are generated by the neutrino mass
matrix, then neutrino mixing parameters are expressed via elements
of the neutrino mass matrix.
\par
The mass lagrangian of two neutrinos ($\nu_e, \nu_\mu$) has the
following form (for simplification the case of two neutrinos is
considered):
$$
\begin{array}{c}{\cal L}_{M} = - \frac{1}{2} \left[m_{\nu_e}
\bar \nu_e \nu_e + m_{\nu_\mu} \bar \nu_{\mu} \nu_{\mu } +
m_{\nu_e \nu_{\mu }}(\bar \nu_e \nu_{\mu } + \bar \nu_{\mu }
\nu _e) \right] \equiv \\
\equiv  - \frac{1}{2} (\bar \nu_e, \bar \nu_\mu)
\left(\begin{array}{cc} m_{\nu_e} & m_{\nu_e \nu_{\mu }} \\
m_{\nu_{\mu} \nu_e} & m_{\nu_\mu} \end{array} \right)
\left(\begin{array}{c} \nu_e \\ \nu_{\mu } \end{array} \right)
\end{array} ,
\eqno(1)
$$
which is diagonalized by rotation on the angle $\theta$ and then
this lagrangian (1) transforms into the following one (see ref. in
[5]):
$$
{\cal L}_{M} = - \frac{1}{2} \left[ m_{1} \bar \nu_{1} \nu_{1} +
m_{2} \bar \nu_{2} \nu_{2} \right]  , \eqno(2)
$$
where
$$
m_{1, 2} = {1\over 2} \left[ (m_{\nu_e} + m_{\nu_\mu}) \pm
\left((m_{\nu_e} - m_{\nu_\mu})^2 + 4 m^{2}_{\nu_\mu \nu_e}
\right)^{1/2} \right] ,
$$
\par
\noindent and angle $\theta $ is determined by the following
expression:
$$
tg (2 \theta)  = \frac{2 m_{\nu_e \nu_\mu}} {(m_{\nu_\mu} -
m_{\nu_e})} , \eqno(3)
$$
$$
\begin{array}{c}
\nu_e = cos \theta  \nu_{1} + sin \theta \nu_{2}  ,         \\
\nu _{\mu } = - sin \theta  \nu_{1} + cos \theta  \nu_{2} .
\end{array}
\eqno(4)
$$
Then $\nu_e, \nu_\mu$ masses are:
$$
m_{\nu_e} = m_1 cos^2 \theta + m_2 sin^2 \theta ,
$$
$$
m_{\nu_\mu} = m_1 sin^2 \theta + m_2 cos^2 \theta , \eqno(5)
$$
in contrast to the primary supposition that $\nu_e, \nu_\mu,
\mu_\tau$ neutrinos have no definite masses.
\par
The probability of $\nu_e \to \nu_e$ is given by the following
expression:
$$
P(\nu_e \rightarrow \nu_e) = 1 -  \sin^{2}(2 \theta) sin^2
((m^{2}_{2} - m^{2}_{1}) / 2p) t , \eqno(6)
$$
where
$$
sin \theta= \frac{1}{\sqrt{2}} \left[ 1 - \frac{| m_{\nu_\mu} -
m_{\nu_e} |}{\sqrt{(m_{\nu_\mu} - m_{\nu_e})^2 + (2 m_{\nu_e
\nu_\mu})^2}} \right]  , \eqno(7)
$$
or
$$
sin^2 (2 \theta) = \frac{(2m_{\nu_{e} \nu_{\mu}})^2} {(m_{\nu_e} -
m_{\nu_\mu})^2 +(2m_{\nu_e \nu_{\mu}})^2} . \eqno(8)
$$
Then the nondiagonal mass term $m_{\nu_e \nu_\mu}$ of the mass
matrix in (1) can be interpreted as width of $\nu_e
\leftrightarrow \nu_\mu$ transitions [4].
\par
In this standard theory of neutrino oscillations neutrino
oscillations are real even when neutrino masses are different
therefore the law of energy momentum conservation is violated. In
the corrected theory of neutrino oscillations [6] the law of
energy momentum conservation is fulfilled and neutrino
oscillations are virtual if neutrino masses are different and real
if neutrino masses are equal.
\par
It is necessary to note that in physics all the processes are
realized through dynamics. Unfortunately, in the above considered
mass mixings scheme the dynamics is absent. Probably, this is  an
indication of the fact that this scheme is incomplete one, i.e.,
this scheme requires a physical substantiation. Below we consider
neutrino oscillations which appear in the scheme of charge (couple
constant) mixings, i.e. by using dynamics [4].

\section{Theory of Neutrino Oscillations in the Framework of Charge Mixings Scheme}

\par
At first we consider a case of two neutrino mixings (oscillations)
and then we consider the case of three neutrino oscillations.

\subsection{The Case of Two Neutrino Mixings (Oscillations)
in the Charge Mixings Scheme}

\par
In this scheme (or mechanism) the neutrino mixings or transitions
can be realized by mixings of the neutrino fields in analogy with
the vector dominance model ($\gamma-\rho^o$ and $Z^o-\gamma$
mixings), the way it takes place in the particle physics. Then, in
the case of two neutrinos, we have
$$
\nu_1 = \cos\, \theta \nu_e - \sin\, \theta \nu_\mu , \eqno(9)
$$
$$
\nu_2 = \sin\, \theta \nu_e + \cos\, \theta \nu_\mu .
$$
The charged current in the standard model of weak interactions for
two lepton families has the following form:
$$
j^\alpha  = \left(\begin{array}{cc} \bar e \bar \mu
\end{array}\right)_L \gamma^\alpha V \left(\begin{array}{c} \nu_e \\
\nu_\mu \end{array} \right)_L ,
$$
$$
V = \left(\begin{array}{cc} \cos \,\theta & -\sin\, \theta \\
\sin \,\theta & \cos\, \theta \end{array}\right) , \eqno(10)
$$
and then the interaction Lagrangian is
$$
{\cal L} = \frac{g}{\sqrt{2}} j^\alpha W^{+}_\alpha  +\rm  h.c.
\eqno(11)
$$
and
$$
\begin{array}{c}
\nu_e = \cos\, \theta  \nu_{1} + \sin\, \theta \nu_{2}           \\
\nu _{\mu } = - \sin\, \theta  \nu_{1} + \cos\, \theta  \nu_{2} .
\end{array}
\eqno(12)
$$
The lagrangian (10)-(11) can be rewritten in the following form:
$$
{\cal L} = \frac{g}{\sqrt{2}} j^\alpha W^{+}_\alpha  +\rm  h.c. ,
\eqno(13)
$$
where $j^\alpha$ is
$$
j^\alpha  = \left(\begin{array}{cc} \bar e \bar \mu
\end{array}\right)_L \gamma^\alpha  \left(\begin{array}{c} \nu_1 \\
\nu_2 \end{array} \right)_L .
$$
And the mass matrix is
$$
\left(\begin{array}{cc} m_1& 0 \\ 0 & m_2
\end{array} \right) .
$$
In this case the neutrino oscillations cannot take place, and even
if neutrino oscillations take place, then there must be $\nu_1
\leftrightarrow \nu_2$ neutrino oscillations but not $\nu_e
\leftrightarrow \nu_\mu$ oscillations.
\par
In this point some questions arise. Where have we taken $\nu_e,
\nu_\mu$ neutrinos if in the weak interactions, given by
expression (13),  $\nu_1, \nu_2$ neutrinos are produced? From the
all existent accelerator experiments very well know that in the
weak interactions $\nu_e, \nu_\mu$ neutrinos are produced and that
the $l_{\nu_e}, l_{\nu_\mu}$ lepton numbers are well conserved
ones. Obviously we must solve this problem. So, $\nu_1, \nu_2$
neutrinos are eigenstates of the weak interactions when we take
mixing matrix $V$ into account and $\nu_e, \nu_\mu$ neutrinos are
eigenstates of the weak interactions with $W, Z^o$ boson
exchanges. Then we have to rewrite the lagrangian of the weak
interaction in the correct form to describe neutrino productions
and oscillations correctly. Then
$$
{\cal L} = \frac{g}{\sqrt{2}} j^\alpha W^{+}_\alpha  +\rm  h.c.
\eqno(14)
$$
where $j^\alpha$ is
$$
j^\alpha  = \left(\begin{array}{cc} \bar e \bar \mu
\end{array}\right)_L \gamma^\alpha  \left(\begin{array}{c} \nu_e \\
\nu_\mu \end{array} \right)_L ,
$$
How are the lepton numbers violated? It is necessary to suppose
that after $\nu_e, \nu_\mu$ production the violation of lepton
numbers takes place, i.e.,
$$
\left(\begin{array}{c} \nu_e \\
\nu_\mu \end{array} \right)_L =
V \left(\begin{array}{c} \nu_1 \\
\nu_2 \end{array} \right)_L , \quad
V = \left(\begin{array}{cc} \cos \,\theta & -\sin\, \theta \\
\sin \,\theta & \cos\, \theta \end{array}\right) , \eqno(15)
$$
and then $\nu_e, \nu_\mu$ neutrinos become superpositions of
$\nu_1, \nu_2$ neutrinos.
$$
\begin{array}{c}
\nu_e = cos \theta  \nu_{1} + sin \theta \nu_{2}  ,         \\
\nu _{\mu } = - sin \theta  \nu_{1} + cos \theta  \nu_{2} .
\end{array}
\eqno(16)
$$
Taking into account that the charges of $\nu_1, \nu_2$ neutrinos
are $g_1, g_2$, we get
$$
g \cos\, \theta = g_1, \quad g \sin\, \theta = g_2 , \eqno(17)
$$
i.e.
$$
\cos\, \theta = \frac{g_1}{g}, \quad \sin \,\theta =
\frac{g_2}{g}. \eqno(18)
$$
Since $\sin^2\, \theta + \cos^2 \,\theta = 1$, then
$$
g = \sqrt{g_1^2 + g_2^2}
$$
and
$$
\cos\, \theta = \frac{g_1}{\sqrt{g_1^2 + g_2^2}}, \quad \sin\,
\theta = \frac{g_2}{\sqrt{g_1^2 + g_2^2}}. \eqno(19)
$$
\par
Since we suppose that $ g_1 \cong g_2 \cong \frac{g}{\sqrt{2}}$,
then
$$
\cos\, \theta \cong  \sin \,\theta \cong \frac{1}{\sqrt{2}} .
\eqno(20)
$$
\par
In the general case the couple constants $g_1, g_2$ and $g$ can
have no connections and then we obtain only expressions (19).
\par
What happens with the neutrino mass matrix in this case? The
primary neutrino mass matrix has the following diagonal form:
$$
\left(\begin{array}{cc} m_{\nu_e}& 0 \\ 0 & m_{\nu_\mu}
\end{array} \right) , \eqno(21)
$$
since in the weak interactions (with $W, Z^o$ bosons) the lepton
numbers are conserved and then $\nu_e, \nu_\mu$ are eigenstates of
these interactions.
\par
It is interesting to note that the same situation takes place in
the quark sector when we consider $K^o, \bar K^o$ oscillations. In
the strong interactions only $d, s, b$ quarks are produced and the
aroma numbers are well conserved in these interactions, i.e.,
these states are eigenstates of the strong interactions. Then
oscillations appear while at violating the aroma numbers by the
weak interactions with the Cabibbo-Kobayashi-Maskawa matrices.
\par
Then due to the presence of terms violating the lepton numbers,
the nondiagonal terms appear in this matrix and then this mass
matrix is transformed into the following nondiagonal matrix (the
case when $CP$ is conserved ):
$$
\left(\begin{array}{cc}m_{\nu_e} & m_{\nu_e \nu_\mu} \\ m_{\nu_\mu
\nu_e} & m_{\nu_\mu} \end{array} \right) , \eqno(22)
$$
then the masses lagrangian of neutrinos takes the following form:
$$
\begin{array}{c}{\cal L}_{M} = - \frac{1}{2} \left[m_{\nu_e}
\bar \nu_e \nu_e + m_{\nu_\mu} \bar \nu_{\mu} \nu_{\mu } +
m_{\nu_e \nu_{\mu }}(\bar \nu_e \nu_{\mu } + \bar \nu_{\mu }
\nu _e) \right] \equiv \\
\equiv  - \frac{1}{2} (\bar \nu_e, \bar \nu_\mu)
\left(\begin{array}{cc} m_{\nu_e} & m_{\nu_e \nu_{\mu }} \\
m_{\nu_{\mu} \nu_e} & m_{\nu_\mu} \end{array} \right)
\left(\begin{array}{c} \nu_e \\ \nu_{\mu } \end{array} \right)
\end{array} .
\eqno(23)
$$
Masses lagrangian of the new states obtained by diagonalizing of
this matrix while rotating on angle $\theta$, has the following
form (these states are namely the same weak interactions states
considered above):
$$
{\cal L}_{M} = - \frac{1}{2} (\bar \nu_e, \bar \nu_\mu) V^{-1}
\left(\begin{array}{cc} m_{\nu_1} & 0 \\
0 & m_2 \end{array} \right) V \left(\begin{array}{c} \nu_e
\\ \nu_{\mu }
\end{array} \right) =
$$
$$
- \frac{1}{2} (\bar \nu_e, \bar \nu_\mu) ) \left(\begin{array}{cc}
m_1 cos^2 \theta + m_2 sin^2 \theta & (m_2 - m_1) cos \theta  sin
\theta \\ (m_2 - m_1) cos \theta sin \theta & m_1 sin^2 \theta  +
m_2 cos^2 \theta \end{array} \right) \left(\begin{array}{c} \nu_e
\\ \nu_{\mu } \end{array} \right) =
$$
$$
- \frac{1}{2} \left[ m_{1} \bar \nu_{1} \nu_{1} + m_{2} \bar
\nu_{2} \nu_{2} \right]  , \eqno(24)
$$
where $\nu_1, \nu_2$ are eigenstates and $m_1, m_2$ are their
eigenmasses and from expressions (23), (24) we obtain
$$
\begin{array}{c}
\nu_1 = cos \theta  \nu_e - sin \theta \nu_\mu  ,  \\
\nu_2 = sin \theta  \nu_e + cos \theta  \nu_\mu ,
\end{array}
\eqno(25)
$$
$$
m_{\nu_e} = m_1 cos^2 \theta + m_2 sin^2 \theta ,
$$
$$
m_{\nu_\mu} = m_1 sin^2 \theta + m_2 cos^2 \theta , \eqno(26)
$$
$$
m_{\nu_e \nu_\mu} = (m_2 - m_1) cos\theta sin \theta ,
$$
or
$$
m_1 = \frac{(m_{\nu_e} cos^2 \theta - m_{\nu_\mu} sin^2 \theta)}
{(cos^2 \theta - sin^2 \theta)} ,
$$
$$
m_2 = \frac{(m_{\nu_e} sin^2 \theta - m_{\nu_\mu} cos^2 \theta)}
{(cos^2 \theta - sin^2 \theta)} , \eqno(27)
$$
where $sin \theta, cos \theta$ are given by expressions (19)
\par
Then $\Delta m^2 = m_2^2 - m_1^2$ is
$$
\Delta m^2 = \frac{(m_{\nu_\mu}^2 cos^4 \theta - m_{\nu_e} sin^4
\theta)}{(cos^2 \theta - sin^2 \theta)^2}  . \eqno(28)
$$
\par
The expression for time evolution of $\nu _{1}, \nu _{2}$
neutrinos (see exp. (25)-(27) with masses $m_{1}$ and $m_{2}$ is
\par
$$
\nu _{1}(t) = e^{-i E_1 t} \nu _{1}(0),  \qquad \nu _{2}(t) =
e^{-i E_2 t} \nu _{2}(0) , \eqno(29)
$$
where
$$
E^2_{k} = (p^{2} + m^2_{k}), \quad k = 1, 2 .
$$
\par
If neutrinos are propagating without interactions, then
\par
$$
\begin{array}{c}
\nu_e(t) = cos \theta e^{-i E_1 t} \nu_{1}(0) + sin \theta
e^{-i E_2 t} \nu_{2}(0) , \\
\nu_{\mu }(t) = - sin \theta e^{-i E_1 t} \nu_{1}(0) + cos \theta
e^{-i E_2 t} \nu_{2}(0) .
\end{array}
\eqno(30)
$$
\noindent Using the expression for $\nu _{1}$ and $\nu _{2}$  from
(25), and putting it into (20), one can get the following
expression:
$$
\nu_e (t) = \left[e^{-i E_1 t} cos^{2} \theta + e^{-i E_2 t}
sin^{2} \theta \right] \nu _e (0) +
$$
$$
+ \left[e^{-i E_1 t} - e^{-i E_2 t} \right] sin \theta \cos \theta
\nu_{\mu }(0) , \eqno(31)
$$
$$
\nu_{\mu }(t) = \left[e^{-i E_1 t} sin^{2} \theta + e^{-i E_2 t}
cos^{2} \theta \right] \nu_{\mu}(0)  +
$$
$$
+ \left[e^{-i E_1 t} - e^{-i E_2 t} \right] sin\theta cos \theta
\nu_e (0) .
$$
\par
The probability that neutrino $\nu_e$ produced at time $t = 0$,
will be transformed into $\nu_{\mu}$ at time $t$, is an absolute
value of amplitude $\nu_{\mu}(0)$ in (31) squared, i. e.
\par
$$
P(\nu_e \rightarrow \nu_{\mu}) = \mid(\nu_{\mu}(0) \cdot \nu_e(t))
\mid^2 =
$$
$$
 = {1\over 2} \sin^{2} (2 \theta) \left[1 - cos ((m^{2}_{2} - m^{2}_{1}) / 2p)
t \right] , \eqno(32)
$$
\noindent where it is supposed that $p \gg  m_{1}, m_{2}$ and
$E_{k} \simeq p + m^{2}_{k} / 2p$.
\par
The expression (32) presents the probability of neutrino flavor
oscillations. The angle $\theta$ (mixing angle) characterizes the
value of mixing. The probability $P(\nu_e \rightarrow  \nu_{\mu})$
is a periodical function of distances where the period is
determined by the following expression:
$$
L_{o} = 2\pi  {2p \over {\mid m^{2}_{2} - m^{2}_{1} \mid}} .
\eqno(33)
$$
\par
And probability $P(\nu _e \rightarrow  \nu _e)$ that the neutrino
$\nu_e$ produced at time $t = 0$ is preserved as $\nu_e$ neutrino
at time $t$, is given by the absolute value of the amplitude of
$\nu_e(0)$  in (31) squared. Since the states in (31) are
normalized states, then
$$
P(\nu_e \rightarrow  \nu_e) + P(\nu_e \rightarrow \nu_{\mu}) = 1 .
\eqno(34)
$$
\par
So, we see that flavor oscillations caused by nondiagonality of
the neutrinos mass matrix violate the law of the $-\ell_e$ and
$\ell_{\mu}$ lepton number conservations. However in this case, as
one can see from exp. (34), the full lepton numbers $\ell  =
\ell_e + \ell_{\mu}$ are conserved.
\par
It is necessary to stress that neutrino oscillations in this
scheme (mechanism) are virtual if neutrino masses are different
and real if neutrino masses are equal and these oscillations are
preserved within the uncertainty relations.  \\

\subsection{The Case of Three Neutrino Mixings (Oscillations)
in the Charge Mixings Scheme}

\par
In the case of three neutrinos we can choose parameterization of
the mixing matrix $V$ in the form proposed by Maiani [12]:
$$
{V = \left( \begin{array} {ccc}1& 0 & 0 \\
0 & c_{\gamma} & s_{\gamma} \\ 0 & -s_{\gamma} & c_{\gamma} \\
\end{array} \right) \left( \begin{array}{ccc} c_{\beta} & 0 &
s_{\beta} \\ 0 & 1 & 0 \\ -s_{\beta} & 0 & c_{\beta}
\end{array} \right) \left( \begin{array}{ccc} c_{\theta} &
s_{\theta} & 0 \\ -s_{\theta} & c_{\theta} & 0 \\ 0 & 0 & 1
\end{array}\right)} , \eqno(35)
$$
$$
c_{e \mu} = \cos\, {\theta } \quad s_{e \mu} =\sin\,{\theta}  ,
 \quad c^2_{e \mu} + s^2_{e \mu} = 1 ;
$$
$$
c_{e \tau} = \cos\, {\beta }, \quad s_{e \tau} =\sin\,{\beta} ,
\quad c^2_{e \tau} + s^2_{e \tau} = 1 ; \eqno(36)
$$
$$
c_{\mu \tau} = \cos\, {\gamma} , \quad s_{\mu \tau}
=\sin\,{\gamma} , \quad c^2_{\mu \tau} + s^2_{\mu \tau} = 1 .
$$
\par
It is not difficult to come to consideration of the case of three
neutrino types $\nu_e, \nu_{\mu}, \nu_\tau$.
\par
For the first and second families (at $\nu_e, \nu_\mu$ neutrino
oscillations) we get
$$
\cos\, \theta = \cos\, \theta_{\nu_e \nu_\mu} =
\frac{g_1}{\sqrt{g_1^2 + g_2^2}} ,
$$
$$
sin (2 \theta) = \frac{2 g_1 g_2}{g_1^2 + g_2^2} . \eqno(37)
$$
Then the probability of $\nu_e \to \nu_e$ is given by the
following expression:
$$
P(\nu_e \rightarrow \nu_e) = 1 - \sin^{2} (2 \theta) sin^2 (\pi t
(m^{2}_{2} - m^{2}_{1}) / 2p_{\nu_e})  , \eqno(38)
$$
In the case $g_1 \cong g_2$
$$
\sin\, \theta_{\nu_e \nu_\mu} \cong \cos\, \theta_{\nu_e \nu_\mu}
\cong \frac{1}{\sqrt{2}} . \eqno(39)
$$
For the first and third families (at $\nu_e, \nu_\tau$ neutrino
oscillations) we get
$$
\cos\, \beta = \cos\, \beta_{\nu_e \nu_\tau} =
\frac{g_1}{\sqrt{g_1^2 + g_3^2}} ,
$$
$$
sin (2 \beta) = \frac{2 g_1 g_3}{g_1^2 + g_3^2} . \eqno(40)
$$
Then the probability of $\nu_e \to \nu_e$ is given by the
following expression:
$$
P(\nu_e \rightarrow \nu_e) = 1 - \sin^{2}(2 \beta) sin^2 (\pi t
(m^{2}_{3} - m^{2}_{1}) / 2p_{\nu_e})  , \eqno(41)
$$
In the case $g_1 \cong g_3$
$$
\cos\, \beta_{\nu_e \nu_\tau} \cong  \sin\, \beta_{\nu_e \nu_\tau}
\cong \frac{1}{\sqrt{2}} . \eqno(43)
$$
For the second and third families (at $\nu_\nu, \nu_\tau$ neutrino
oscillations) we get
$$
\cos\, \gamma = \cos\, \gamma_{\nu_\mu \nu_\tau} =
\frac{g_2}{\sqrt{g_2^2 + g_3^2}} ,
$$
$$
sin (2 \gamma) = \frac{2 g_2 g_3}{g_2^2 + g_3^2} . \eqno(44)
$$
Then the probability of $\nu_\mu \to \nu_\mu$ is given by the
following expression:
$$
P(\nu_\mu \rightarrow \nu_\mu) = 1 - \sin^{2}(2 \gamma) sin^2 (\pi
t (m^{2}_{3} - m^{2}_{2}) / 2p_{\nu_\mu})  , \eqno(45)
$$
In the case $g_2 \cong g_3$
$$
\cos\, \gamma_{\nu_\mu \nu_\tau} \cong  \sin\, \gamma_{\nu_\mu
\nu_\tau} \cong \frac{1}{\sqrt{2}} . \eqno(46)
$$
\par
So the neutrino mixings (oscillations) appear due to the fact that
at neutrino production the eigenstates of the weak interactions
(i.e. $\nu_e, \nu_\mu, \nu_\tau$ neutrino states) are generated
 but not the eigenstates of the weak interaction violating
lepton numbers (i.e. $\nu_1, \nu_2, \nu_3$ neutrino states). And
when neutrinos are passing through vacuum they are converted into
superpositions of $\nu_1, \nu_2, \nu_3$ neutrinos and through
these intermediate states the oscillations (transitions) between
$\nu_e, \nu_\mu, \nu_\tau$ neutrinos are realized.

\section{Conclusion}

It is necessary to note that in physics all the processes are
realized through dynamics. Unfortunately, in the standard theory
of neutrino oscillations based on the masses mixings scheme
(mechanism), the dynamics is absent. Probably, this is an
indication of the fact that this scheme is incomplete one, i.e.,
this scheme requires a physical substantiation.
\par
In this work neutrino oscillations generated by the weak
interaction couple constant (charge) mixings where considered [4]
(as it takes place in the model of vector dominance [7] or in the
electroweak interactions model [8] at vector boson mixings).
Expressions for angle mixings and lengths of oscillations were
obtained. The expressions of probabilities for three neutrino
oscillations were given. Neutrino oscillations in this scheme
(mechanism) are virtual if neutrino masses are
different and real--if neutrino masses are equal. \\

\newpage
\par
{\bf References}\\

\par
\noindent 1. Pontecorvo B. M., Soviet Journ. JETP, 1957, v. 33,
p.549;
\par
JETP, 1958,  v.34, p.247.
\par
\noindent 2. Maki Z. et al., Prog.Theor. Phys., 1962, vol.28,
p.870.
\par
\noindent 3. Pontecorvo B. M., Soviet Journ. JETP, 1967, v. 53,
p.1717.
\par
\noindent 4. Beshtoev Kh. M., JINR Communication E2-2004-58,
Dubna, 2004;
\par
hep-ph/0406124, 2004.

\par
\noindent 5. Bilenky S.M., Pontecorvo B.M., Phys. Rep.,
C41(1978)225;
\par
Boehm F., Vogel P., Physics of Massive Neutrinos: Cambridge
\par
Univ. Press, 1987, p.27, p.121;
\par
Bilenky S.M., Petcov S.T., Rev. of Mod.  Phys., 1977, v.59,
\par
p.631.
\par
 Gribov V., Pontecorvo B.M., Phys. Lett. B, 1969, vol.28,
p.493.
\par
\noindent 6. Beshtoev Kh.M., JINR Commun. E2-92-318. Dubna, 1992;
\par
Beshtoev Kh.M., Proc. of 27th Intern. Cosmic Ray Conf. Germany,
\par
2001, V.4, P.1186;
\par
Beshtoev Kh.M., Proc. of 28th Intern. Cosmic Ray Conf. Japan,
2003,
\par
V.1, P.1503.
\par
Beshtoev Kh.\,M., JINR Commun. E2-2004-58. Dubna, 2004.
\par
\noindent 7. Sakurai J.J., Currents and Mesons. The Univ. of
Chicago
\par
Press, 1967;
\par
Beshtoev Kh.M., Preprint INR of Academy of Sciences of USSR.
P-217.
\par
Moscow, 1981.

\par
\noindent 8. Glashow S. L., Nucl. Phys. 1961. V.22, P.579;
\par
Weinberg S., Phys. Rev. Lett. 1967. V.19, P.1264;
\par
Salam A., Proc. of the 8th Nobel Symp. Ed. N. Svarthholm. Almgvist
\par
and Wiksell, Stockholm, 1968., P.367.

\end{document}